\documentclass[letterpaper,twocolumn,10pt]{article}
\usepackage{usenix-2020-09}
\usepackage{tikz}
\usepackage{amsmath}

\usepackage{booktabs} 
\usepackage{enumerate}
\usepackage{epsfig}
\usepackage{graphicx}
\usepackage{epstopdf}
\usepackage{amsmath}
\usepackage[linesnumbered,ruled]{algorithm2e}
\usepackage{url}
\usepackage{breakurl}
\usepackage{listings}
\usepackage{subfigure}
\usepackage{rotating}
\usepackage{lscape}
\usepackage{verbatim}
\usepackage{hhline}
\usepackage{textcomp,booktabs}
\usepackage{xcolor}
\usepackage{paralist}
\usepackage{hyperref}
\usepackage{grffile}
\usepackage{mathtools}
\usepackage{amsthm}
\usepackage{flushend}
\usepackage{hyphenat}
\usepackage[autostyle,threshold=0]{csquotes}
\usepackage{tabularx}
\usepackage{ulem}
\usepackage{CJKutf8}
\usepackage{amssymb}
\usepackage{tikz}

\let\OLDthebibliography\thebibliography
\renewcommand\thebibliography[1]{
  \OLDthebibliography{#1}
  \setlength{\parskip}{0pt}
  \setlength{\itemsep}{0pt plus 0.3ex}
}

\begin{document}

\date{}

\title{\Large \bf A Study of China's Censorship and Its Evasion Through the Lens of Online Gaming}

\author{
{\rm Yuzhou Feng}\\
FIU
\and
{\rm Ruyu Zhai}\\
Hangzhou Dianzi University
\and
{\rm Radu Sion}\\
Stony Brook University
\and
{\rm Bogdan Carbunar}\\
FIU
}


\maketitle

\begin{abstract}
For the past 20 years, China has increasingly restricted the access of minors to online games using addiction prevention systems (APSes). At the same time, and through different means, i.e., the Great Firewall of China (GFW), it also restricts general population access to the international Internet. This paper studies how these restrictions impact young online gamers, and their evasion efforts. We present results from surveys ($n$ = 2,415) and semi-structured interviews ($n$ = 35) revealing viable commonly deployed APS evasion techniques and APS vulnerabilities. We conclude that the APS does not work as designed, even against very young online game players, and can act as a censorship evasion training ground for tomorrow's adults, by familiarization with and normalization of general evasion techniques, and desensitization to their dangers. Findings from these studies may further inform developers of censorship-resistant systems about the perceptions and evasion strategies of their prospective users, and help design tools that leverage services and platforms popular among the censored audience.
\end{abstract}

\section{Introduction}
\label{sec:introduction}

Online games, partially or primarily played through the Internet~\cite{AEA06}, are predicted to reach 1 billion users by 2024, with revenues to exceed 17 billion USD~\cite{OnlineGamingPredictions}. China is the highest revenue generator in the online game space, with more gamers than the US, Japan, Germany, France and the UK combined~\cite{GameBooster}. In particular, 108 million minors in China are online gamers~\cite{CNOnlineGamers20}.

To address concerns about the well-being of minor gamers~\cite{FB96, HBSSKZW20, CHB18, PFFHMMABGD16, KG12, J14}, various branches of the Chinese government have been involved in defining {\it anti-addiction} policies (AAPs), and mandating {\it addiction prevention system} (APS) implementations by online games and gaming platforms. These policies have set increasingly stringent limitations on gaming activities for minors. For instance, while before September 2021, Chinese minors could play 9-16 hours per game platform per week, they can now only play during three 1-hour slots per week (8-9pm, Fri - Sun) across all compliant platforms. The policies were recently declared successful in eliminating gaming addiction in minors~\cite{AddictionSolved}.

This paper seizes a unique opportunity to explore and push the boundaries of our knowledge on the Chinese digital restrictions landscape, primarily through the lens of online gamers. Such knowledge can inform developers of censorship-resistant systems (CRSes) about the perceptions and evasion strategies of their prospective users.

We build on the relationship between APSes implemented by compliant online games, and the great firewall (GFW): APSes restrict gaming access to vulnerable minors, while the GFW restricts Internet access to a population considered vulnerable to Western influence. Our quest relies on guidance from young Chinese Internet users with educated backgrounds, and aims to answer the following questions:

\begin{compactitem}

\item
{\bf RQ1}: Do anti-addiction policies really affect educated Chinese gamers and how?

\item
{\bf RQ2}: Were minors with educated backgrounds motivated and able to evade APSes? (a) How? (b) Do their strategies and techniques still work today?

\item
{\bf RQ3}: Does the GFW affect such online gamers? (a) Have they attempted to evade the GFW? (b) What challenges are they facing in doing so?

\end{compactitem}

To answer these questions, we first conducted a survey ($n$ = 2,415) with young Chinese users with educated backgrounds who played online games while minors, to probe their exposure to APSes. We then conducted interviews with a subset ($n$ = 35) of the survey respondents who were online gamers, to study their perceptions of anti-addiction policies and the GFW, and their evasion efforts. The findings include:

\begin{figure}
\centering
\includegraphics[width=0.95\columnwidth]{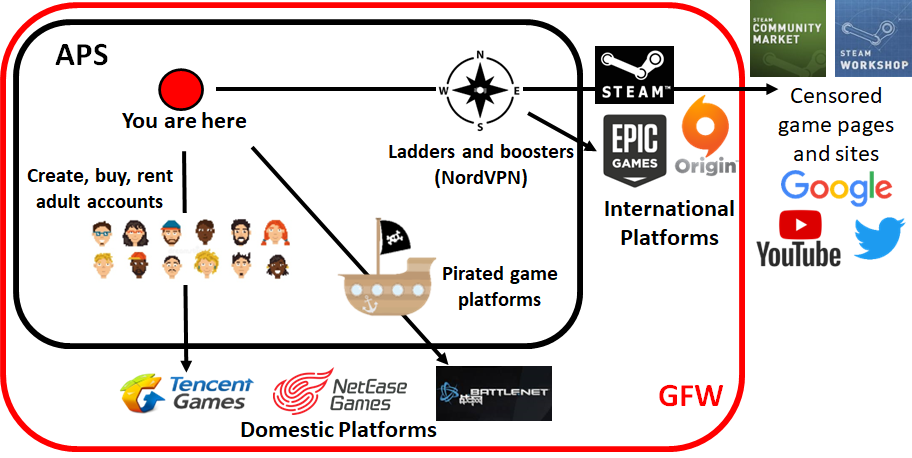}
\caption{Map of participant-revealed strategies to evade addiction prevention systems (APSes). Several strategies continue to work for the latest APSes.}
\label{fig:map:evasion}
\vspace{-5pt}
\end{figure}

(1) AAP Impact: Many survey participants were affected by anti-addiction policies while minors, and evaded APSes. Survey respondents potentially affected by more restrictive AAPs were more likely to declare being affected by these policies, to have evaded APSes, and to have used GFW evasion tools and gaming accelerators~\cite{UUBooster, TencentBooster}.

(2) APS Evasion: Minors employ diverse strategies to evade gaming restrictions, which include acquiring adult accounts, using international gaming platforms or domestic servers hosting pirated games, and using VPNs and gaming accelerators to boost access to foreign games and access censored sites. Figure~\ref{fig:map:evasion} outlines a map of participant-revealed APS-evasion strategies, and points out vulnerabilities in current APSes, that may be used to develop future CRSes.

(3) GFW Impact and Evasion Insights: Many interview participants were affected by the GFW in game playing or online browsing activities. While some GFW-evasion tools fail frequently, others can be easily found online. Some participants have privacy concerns related to their use of evasion tools, and some reported exposure to undesirable content when searching and installing such tools.

We further introduce the conjecture, to be fully evaluated in future work, that evasion needs for trivial game-playing activities, normalize the need to evade authority, desensitize users to the dangers of GFW evasion, and may even train young Chinese users to evade the GFW.

Findings from these studies may help CRS developers design censorship-evading tools that leverage technologies, services and platforms popular among their prospective users.

\section{Related Work}

This section discusses related work on gaming anti-addiction policies and the great firewall of China. It reports gaps in knowledge concerning the impact of gaming and Internet restrictions on minors, the techniques and strategies used to evade them, and relationships between evading gaming and Internet restrictions.

\noindent
{\bf Impact of Video Games}.
Significant previous work has focused on the effects of video gaming. Part of previous research has focused on negative effects that include violence~\cite{PSH18}, gaming disorder~\cite{KDPDKKB19}, and the effects of increased gaming time for young people on their social competence~\cite{HBSSKZW20}, conduct, peer relations~\cite{PFFHMMABGD16} and mental health~\cite{JTYWPJ18}. The negative impact on peer relations was observed only in children playing video games 9 hours or more per week~\cite{PFFHMMABGD16}, while the social competence study recorded a maximum playing time of under 10 hours. The third generation of Chinese anti-addiction policies reduced the play time for Chinese minors from 10.5 to 3 hours per week. Other research, albeit not focused on minors, has shown that video gaming can provide benefits in spatial cognition~\cite{SF10, F07}, working memory and planning~\cite{CWZH13}, social organization and communication~\cite{F10, ESA21, S13}, and mental health and well-being~\cite{JSJKC14}. Other work has further argued that games may simultaneously have both positive and negative effects~\cite{F10, S13}.

In particular, gaming disorder, including in-game gambling, is a global problem, added by the World Health Organization (WHO) to its International Classification of Diseases-11 (ICD-11). This paper focuses on the effects of restrictions imposed by anti-addiction policies and not on the effects of online gaming, including gaming disorder.

\noindent
{\bf Gaming Addiction Prevention}.
Efforts to implement and evaluate online gaming addiction prevention measures include promoting critical reflection and recognizing gaming problems~\cite{KD17}, school-based programs~\cite{JP10}, and parent-based programs~\cite{ACC19}. Previous work suggests that restrictions set by parents are not effective in reducing pathological symptoms of their children's video-gaming behaviors~\cite{CSLGK15}. Indeed, several of our interview participants would prefer parental control to the APSes of online games. Some evaded APSes with parental complicity, and others surreptitiously exploited parental trust for evasion purposes.

\noindent
{\bf The Great Firewall, Censorship and Perceptions}.
Internet censorship in China focuses on both surveillance and access restrictions for content hosted externally, and surveillance and censorship of internal communications. Censorship occurs through the Great Firewall (GFW) of China, which employs techniques that include DNS-based censorship of entire sites~\cite{LWM07, FDW16}, application-level content filtering~\cite{CCKMSTW13} and keyword filtering~\cite{KRC17, WBC21}.

Previous studies found that many Chinese Internet users may conform to government ideologies, are not interested in political matters~\cite{W14, WM15}, and are not aware of Internet censorship or have significant support for it, even if they bypass it~\cite{KSN17, WM15}. However, while many Chinese users engage with government platforms for reasons that include compliance and patriotism~\cite{LX20}, perceived Internet censorship was shown to negatively affect user continuance of mobile government micro-blogging services ~\cite{LYCY18}.

This points to a need for more research to determine support for restrictive policies among the people who experience them. In our queries of support for gaming policies among young people who experienced them, we found polarized and conflicting perceptions. This includes participants who consider these policies too relaxed, after no longer being affected by them, and criticism of the policies among their supporters.


Lu et al.~\cite{LJLNW20} studied perceptions of Chinese WeChat users on misinformation in the context of censorship and astroturfing, and their impact on news consumption practices. Kou et al.~\cite{KKGB17} studied strategies used by Chinese users to navigate information allowed by censorship in social networks, state, and foreign media. Our efforts instead focus on (1) how young people affected by anti-addiction policies were also affected by, and evaded Internet censorship, (2) relationships between techniques used to evade the GFW and the APS, and (3) participant concerns with the use of evasion tools.

\noindent
{\bf Studies of Gambling in China}.
Part of this work is related to studies of (illegal) online gambling in China. Yang et al.~\cite{YDZHLLWDSS19} analyzed the ecosystem of illegal online gambling in China, including their profit chain. Gao et al.~\cite{GWLLXL21} analyzed mobile gambling apps, and proposed techniques to identify new gambling services. Hong et al.~\cite{HGY22} explored the operational pipeline and fraud kill chain of Chinese mobile gambling scams. In contrast to these quantitative analyses of online gambling, part of our qualitative study explores exposure to vulnerabilities by online gamers during APS and GFW evasion efforts. Several interview participants confirmed exposure to ads for illegal services, including online gambling.

\noindent
{\bf Game-Based Evasion}.
Online games have been proposed as a cover for censorship circumvention and even plausibly deniable communications~\cite{VK15,HNPJ16}. For instance, Rook~\cite{VK15} hides data into gaming packets, while Castle~\cite{HNPJ16} encodes data into commands in RTS games. Our work suggests further potential for gaming accounts to surreptitiously store and communicate information.

\section{Background}
\label{sec:background}

\begin{figure}
\centering
\includegraphics[width=\columnwidth]{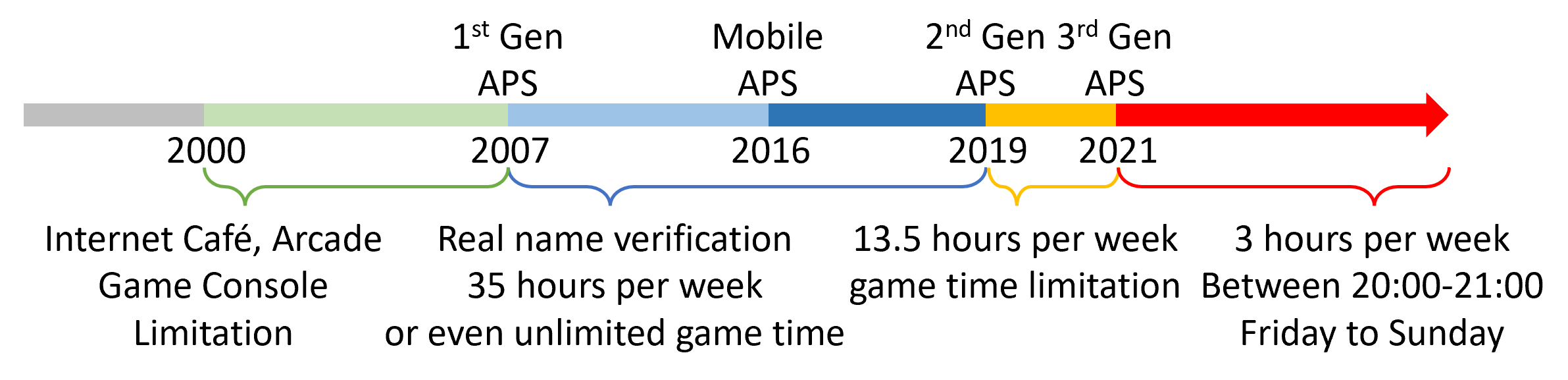}
\vspace{-10pt}
\caption{Timeline of Chinese anti-addiction policies.}
\vspace{-5pt}
\label{fig:aap:timeline}
\end{figure}

The Chinese Online Game Addiction Prevention Project (\begin{CJK*}{UTF8}{gbsn}未成年人网络游戏防沉迷系统\end{CJK*}) is a comprehensive effort where the government issues policies, to be implemented by online game providers, parents, and schools, and to be enforced by legal supervisory departments~\cite{ZJCH12}. Figure~\ref{fig:aps:components} outlines key components of anti-addiction policies (AAPs) and the corresponding components of the addiction prevention systems (APSes) implemented by online games. In the following we first present the history of anti-addiction policies, then describe main APS components.

\subsection{History of Anti-Addiction Policies}
\label{sec:background:timeline}


The process to define anti-addiction policies in China has started in the year 2000, is constantly evolving, and continues through this day. Figure~\ref{fig:aap:timeline} shows AAP milestones until the time of this study. Section~\ref{sec:discussion:future} discusses more recent changes. The following outlines the chronological evolution of these policies, organized by AAP generation.

\noindent
{\bf Early Days}.
In 2000, the Chinese General Office of the State Council released a notice to stop all the production, sales, and business activities of game consoles~\cite{GOSC00}. In 2002, the government prohibited Internet cafes from admitting minors~\cite{SC02}.
%

\noindent
{\bf First Generation AAP}.
In 2007, the Chinese State Administration of Press, Publication, Radio, Film and Television (SAPPRFT) issued a notice requiring online game operators to implement an anti-addiction policy and real-name verification system, and reduce the value of in-game benefits of minors by half after three hours of continuous play, and to zero after five hours. The SAPPRFT further extended the APS policy to cover mobile games in 2016~\cite{SAPPRFT17}.

\noindent
{\bf Second Generation AAP}.
In 2019, The National Press and Publication Administration (NPPA), replaced SAPPRFT as China’s regulatory authority for online publishing. The NPPA released a second generation anti-addiction notice~\cite{NPPA19} in the same year. It required game providers to prevent access for users that have not completed the real-name registration. It also reduced gaming times for minors to 1.5 hours per day and introduced a gaming curfew for minors (10pm - 8am). Further, it restricted in-game purchases.

\noindent
{\bf Third Generation AAP}.
In 2021, the NPPA released the most restrictive addiction prevention notice~\cite{NPPA21}, asking online game providers to further restrict access for minors to three 1-hour slots per week (Friday-Sunday, 8-9pm)~\cite{CGIGC21}.

\noindent
{\bf Licensing of Online Games}.
In 2009, SAPPRFT/NPPA released a notice~\cite{NPPA09} to clarify that in order to operate in China, any games, including international ones, must be pre-approved, and obtain an Internet publishing license~\cite{PublishGameInChina}. Compliance with the above anti-addiction policies is a prerequisite for licensing. The license application needs to include the game's source code, operating instructions, admin credentials, dialogue, script, descriptive text, song lyrics (all in Chinese), and pictures of main characters~\cite{X12}. While licensing criteria are vague and change over time, forbidden content includes gambling, violence (e.g., blood), sexual content, and also content that harms national security~\cite{X12}. The NPPA limits the number of games licensed each year.

\noindent
{\bf Licensing of International Games}.
The NPPA requires that all international online games partner with a domestic operator. The delegated operator acts on behalf of the publisher to deal with marketing and legal issues in China, including the licensing process. The license is then linked to the domestic operator. This was the case for the recently defunct partnership between Activision Blizzard and China's NetEase platform~\cite{ActivisionNetEase}. All the Activision Blizzard games are now unlicensed in China; the company will need to set up a partnership with another domestic operator and re-apply for licenses.

In summary, the evolution of anti-addiction policies has introduced ever-increasing limitations on all the aspects of game playing experiences of minors, including their ability to open accounts on gaming platforms, the types of games they can access, gaming times, in-game points, and purchases.

\subsection{APS Components}
\label{sec:aps:components}

\begin{figure}[t]
\centering
\includegraphics[width=\columnwidth]{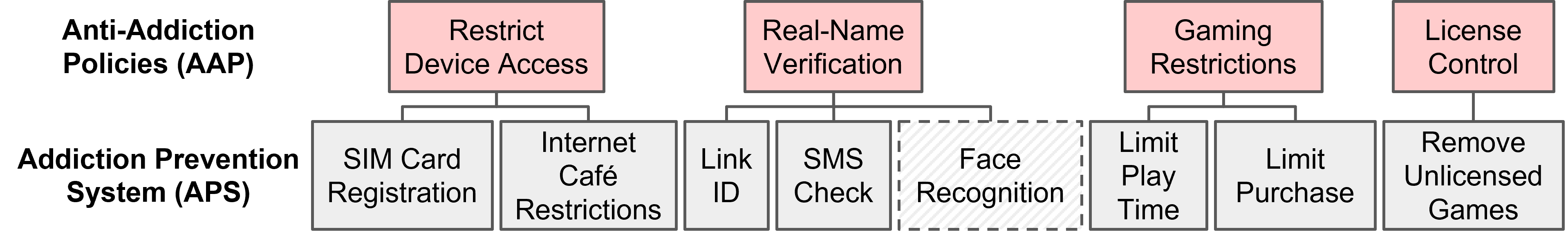}
\vspace{-10pt}
\caption{Key anti-addiction policy (AAP) and addiction prevention system (APS) components. They include device access restrictions for minors, mandates of name and age verification during game platform account registration, gaming restrictions for minors, and bans on unlicensed games.}
\vspace{-5pt}
\label{fig:aps:components}
\end{figure}

This section presents key components of addiction prevention systems implemented by online games and platforms, to be in compliance with anti-addiction policies, see Figure~\ref{fig:aps:components}.

\noindent
{\bf Real-name Verification}.
The real-name verification system requires online gamers to link a Chinese national ID to their accounts. The Chinese ID number consists of a 17-digit body code and a one-digit verification code. The body code consists of a six-digit address code, eight-digit date of birth code, and a three-digit sequence code. Several gaming platforms (e.g., Tencent, Battle.net.cn) further enhanced their real-name verification process with the following two components:

$\bullet$
{\bf SMS-Based Verification}.
Starting in 2017, online game operators in China impose a mobile phone-based ID verification process: Users need to provide their name, national ID number, and mobile phone number to register an account, or continue using their accounts. Users need to confirm the phone number by retrieving an SMS from the phone.

$\bullet$
{\bf SIM Card Registration}.
In 2017, state-owned telecom companies introduced a SIM card registration process, to connect each SIM card in use to a verified user. To register a new SIM, the user needs to link it to her national ID in the SIM provider's database. The registration process includes a facial recognition step; since 2019, this can be done via mobile apps. Each national ID can be associated with at most five SIM cards for each telecom company. Thus, to open accounts on some gaming platforms, users need to link a national ID, and validate ownership of a registered SIM card.

\noindent
{\bf Platform-Specific Components}.
Game providers also develop their own addiction prevention features, not AAP mandated. For instance, in 2018, Tencent's mobile gaming platform implemented facial recognition of its users~\cite{Tencent18}, a feature provided in collaboration with the police~\cite{TencentFacial}. The process validates the user's ID and age, and is triggered by behaviors indicative of minors, e.g., adult account that plays a suspiciously long time at least once a week, or registering a gaming account for a user 60 years old or above~\cite{Tencent18}.

Further, during the second generation APS, Netease developed the NetEase Parent Care Platform~\cite{NetEase18}, which allows parents to monitor and control game hours and purchases.

\subsection{Terminology}

A game is said to be {\it licensed}, if it was reviewed and approved by the NPPA. A gaming platform is {\it compliant} if it implements the NPPA-specified anti-addiction policies ($\S$~\ref{sec:background:timeline}).

We adopt the term GFW {\it ladder} (\begin{CJK*}{UTF8}{gbsn}梯子\end{CJK*}) from study participants, to denote censorship evasion tools that include VPNs and game {\it boosters} (or {\it accelerators})~\cite{UUBooster, TencentBooster}. Game boosters are tools to reduce the latency experienced when playing online games, particularly those hosted on remote, international servers. We further use the expression ``climb over the wall'' (\begin{CJK*}{UTF8}{gbsn}翻墙\end{CJK*}) to denote GFW evasion activities.

\section{Methods}
\label{sec:methods}

The paper presents results from a mixed-methods approach that combines a larger survey with a more fine-tuned qualitative study, based on individual in-depth interviews.

\subsection{Online Survey}
\label{sec:methods:survey}

The survey was designed to study the impact of online gaming anti-addiction policies on young Chinese Internet users, and to identify candidates for interviews ($\S$~\ref{sec:methods:recruitment}). Participants were presented with yes/no assertions that explore their exposure to gaming addiction prevention systems, and gently probe their exposure to evasion technologies, including for the APS and the GFW. The survey includes a single assertion about the use of GFW evasion tools, purposely double-barreled, i.e., probing the use of GFW evasion tools {\it or} game accelerators to access international gaming platforms.

The survey was translated in Chinese by a team member who is a native speaker. Participants needed to first provide their age; data was not collected from participants younger than 18. The survey is included in Appendix~\ref{appendix:survey}.

\subsection{Semi-Structured Interviews}
\label{sec:methods:interview}

The qualitative study investigates online gaming experiences of young Chinese adults (while minors), their perception and evasion of the APS, and exposure to and evasion of the GFW. The interview questions probe the impact of the APS and GFW on online gamers, to understand whether they evaded them, their evasion reasons, strategies, and experiences.

Semi-structured interviews were conducted with qualified and consenting survey participants, see Appendix~\ref{appendix:interview} for questions. Interviews were conducted online by a team member who is a native Chinese speaker, over live chat applications indicated by participants. Audio was recorded with consent.
%
%
Interviews started with a description of the team, the objectives, procedures and risks of the study. Interviews lasted approximately one hour each (M = 41 mins, SD = 19 mins).

\noindent
{\bf Data Analysis}.
Interviews were translated, transcribed, pseudonymized, and securely stored. The English-to-Chinese translation of the interview questions, and the Chinese-to-English translation of participant responses, were performed by the same team member. Translation was followed by back-translation using Google Translate, to ensure that the original meaning was preserved. Due to English vs. Chinese language differences, to ensure comprehension and establish rapport, the translation occasionally deviated from literal translation, including to use technical slang, abbreviations, alternative keywords popular among young Chinese netizens, e.g., ``climbing the wall'', ``ladder'', ``accelerator''. Each transcript was further discussed by a non-speaker author with the translator to eliminate translation ambiguities before analysis.

Each interviewee was assigned an id that corresponds to their order of participation in the initial survey. Transcripts were then analyzed using applied thematic analysis~\cite{GME11, GME12}, by systematically generating and iteratively conceptualizing codes and themes: the data was analyzed, and important pieces, i.e, codes, were identified and grouped into themes.

More specifically, two researchers independently read the first five transcripts, coded responses to each interview question, then organized them into themes through an initial codebook. The researchers met to discuss the themes and codes, and revise the codebook. The researchers repeated this process over batches of 5 - 7 interview transcripts. After each batch, agreement was reached between the codes and themes identified by the researchers. At the completion of this process, the researchers independently applied the identified codebook to all the interview transcripts.

\subsection{Participant Recruitment}
\label{sec:methods:recruitment}

Survey participants were recruited using the snowball sampling method~\cite{BW81}, by first posting the survey on the Chinese crowd-sourcing site wenjuan.com, and obtaining the QR code embedding the link to the job. The researchers distributed the code to friends, colleagues, and contacts, and also posted the link on bulletin board systems of Chinese universities and technical colleges, micro-applications like QQ and WeChat, and chat groups of Chinese online games~\footnote{We did not recruit participants from other regions (e.g., Hong Kong, Macao, Taiwan), where AAPs are not enforced.}.


At the end of the survey, participants were asked if they would like to participate in a one-on-one interview, and provide preferred contact information. Survey respondents who expressed interest (and provided contact information), and who played online games and evaded APSes while minors, were invited to the interview. All interview participants were asked to distribute the survey's link to their contacts.

Survey participants were not paid. Interview participants were paid a rate of 10 CNY (1.5 USD) for every 15 minutes spent in the interview.

In total, 3,427 people answered the survey. Seventy of the respondents, chosen randomly among those who qualified, were invited to the interview. Interviews were conducted with the 35 who responded and consented. Saturation was reached after the 25th interview, i.e., no {\it new} information was subsequently identified, relevant to the research questions of $\S$~\ref{sec:introduction}.

\subsection{Ethical Considerations}
\label{sec:methods:ethical}

The study carefully followed established ethical practices for conducting sensitive research with vulnerable populations, including clearly declaring the researchers' identity and research objective, without following any deception. The procedure was scrutinized and approved by the institutional review board at FIU (IRB-21-0536).

\noindent
{\bf Informed Consent}.
Survey participants were presented with a consent form before starting the survey. Interview participants were again provided with a consent form and were asked to read and indicate consent. The risks were discussed in detail, and participants were given ample opportunity before the start of the interview to evaluate these risks, ask questions, and provide consent. During the interview, participants were further asked if they are comfortable discussing sensitive topics (question I10 for APS evasion and question I15 for GFW evasion, see Appendix~\ref{appendix:interview}) before asking questions.

\noindent
{\bf Data Management}.
During the survey recruitment process the researchers had no access to any personally identifiable information of users. Survey participants were given the opportunity to provide minimal contact information, i.e., either e-mail address, instant messaging account name, or phone number, if interested to participate in a later interview. All contact information from survey participants who did not qualify for the interview, did not reply to the invitation, or did not participate in an interview, was destroyed.


After the interview, each participant provided a preferred payment method. The participant's PII (contact and payment information) was destroyed immediately after the payment process. Answers collected during the interview were first pseudonymized, then were stored on a physically secure Linux server in our university, and accessed through encrypted channels only from the password-protected authors' laptops.

\noindent
{\bf Risks}.
We structured our research to minimize risks to participants, but could not eliminate the risk. As described below, we concluded that the risks to informed participants were minimal and no greater than necessary to our inquiry. Study risks have been disclosed to and discussed with participants prior to commencing the study, see Appendix~\ref{appendix:consent}.

More specifically, participants were told that some questions may be upsetting, and were instructed to skip any questions they do not feel comfortable answering, both before and during the interview (e.g., questions I10 and I15, Appendix~\ref{appendix:interview}). Participants were told that once published, results may be used by other parties, including online game providers and the GFW provider, to try to close any gaps in their defenses. Participants were told they run this risk in the regular interaction with these highly monitored systems anyway.

All participants were over 18 years old, and the anti-addiction policies and systems only apply to minors. Thus, participant risks of discussing APS evasion does not exceed the risk of evading the APS while being minor. Further, the risks for participants who said they evaded the GFW do not exceed the risks already incurred by evading the GFW, subject to extensive GFW monitoring activities. This is because a censor is likely to already have all the necessary information to assert if the participant is evading the GFW through mandatory reports from ISPs, logs of IP addresses of VPNs that they use, etc.

Recall that we deleted payment information immediately after interviews. This reduces the risk that participants could be linked to the interview study. If such a link were made, it would not directly connect participants to their interview answers.

We further note that recruiting Chinese netizens externally, without an easy trust basis mechanism, and asking questions about restriction evasion is challenging. Participation risks were a foremost concern when designing the survey and interview questions. Indeed, the questions impose a trade-off between efforts to collect novel, sensitive insights that provide comprehensive coverage of online restrictions, and participation risks, thus also the willingness of Chinese netizens to participate. Therefore, the survey and interview questions avoided sensitive political or societal topics.

\section{Survey Findings}
\label{sec:survey}

\begin{table}[t]
\centering
\resizebox{\columnwidth}{!}{
\textsf{
\begin{tabular}{l c c c c}
\toprule
Questions   & Total   & Male   & Female   & Undeclared\\
\midrule
Online Gamer when Minor   & 2415   & 556   & 1534   & 325 \\
Affected by APS   & 1786   & 425   & 1105   & 256 \\
{\bf Evaded APS}   & {\bf 674}   & {\bf 202}   & {\bf 374}   & {\bf 98} \\
Played on international Platform   & 361   & 191   & 64   & 106 \\
{\bf Used GFW Ladder or Booster}   & {\bf 279}   & {\bf 201}   & {\bf 78}   & {\bf 0} \\
Understands Foreign Language   & 336   & 150   & 67   &  99 \\
\bottomrule\\
\end{tabular}}}
\caption{Summary of survey answers over the 2,415 respondents that played online games while minors.}
\label{table:survey}
\vspace{-15pt}
\end{table}


The survey received responses from 3,427 users with an educated background. All respondents were between 18 and 78 years old. Data was discarded from participants who (i) were over 37 years old, since they were not affected by (any versions of) the anti-addiction policies, (ii) only answered the first survey question (providing their age), (iii) provided the same answer to all the assertions, in under 40 seconds, and (iv) answered ``No'' to question S2 in $\S$~\ref{appendix:survey} (did not play online games while minors) but answered “Yes” to either S3, S4, S5, S7, or S8 (made in-game purchases, where affected by or evaded the APS, or played international games).

Of the remaining 3,364 survey respondents, 18.46\% are male, 69.56\% are female, and 11.98\% preferred not to answer. The reason for the gender imbalance may be the fact that many participants were recruited from two technical colleges with high female concentrations. 69.82\% of the respondents were university students, 26.87\% were students in technical colleges, 1.8\% were employed, and 0.5\% were unemployed.

More than 71\% (2,415) of the 3,364 survey respondents, played online games when they were minors (under 18). 23\% of these are male, and 63.51\% are female. The rest did not declare their gender. Table~\ref{table:survey} summarizes the survey answers for these participants.

More than 73\% (1,786) of these respondents were affected by the APS when they were minors. Similar percentages among male and female respondents in their own gender group are observed. All these 1,786 respondents were affected by the first generation anti-addiction policies.

Out of the 1,786 respondents affected by the APS when they were underage, 37.7\% (674) successfully evaded the APS. More than 36\% of the male gamers evaded the APS; A relatively large percentage (24.12\%) of female gamers have also evaded the APS.

\begin{figure}
\centering
\includegraphics[width=\columnwidth]{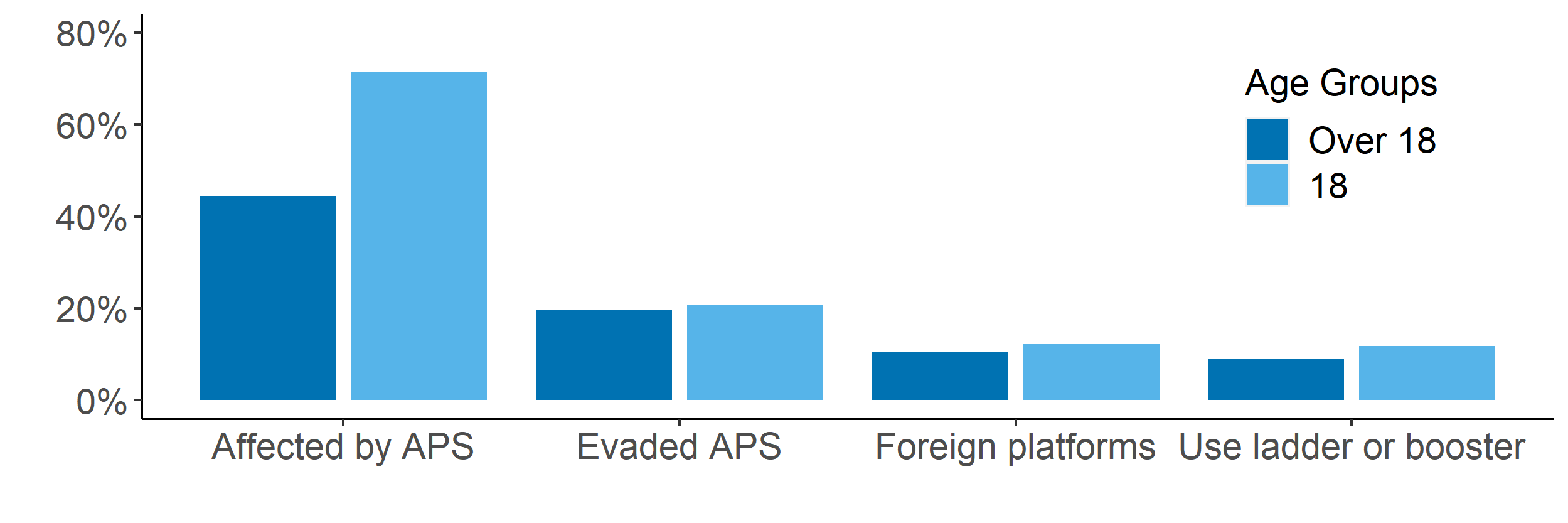}
\vspace{-10pt}
\caption{Impact of the third generation anti-addiction policies. The third generation resulted in a significant increase in the percentage of participants affected by these policies, and smaller increases in the percentage of those that engaged in evasion behaviors.}
\vspace{-5pt}
\label{fig:survey:gen3}
\end{figure}

361 respondents played on international gaming platforms. To access these platforms, 279 of them used either a GFW evasion tool or a gaming accelerator (booster). Gaming accelerators~\cite{UUBooster, TencentBooster}, considered to be ``games-only VPNs''~\cite{ChineseGameMarket, GameBooster}, reduce the latency experienced when playing online games. When compared against male respondents, substantially smaller percentages of females played games on international platforms (4.24\% vs. 34.89\%), and used GFW evasion or game acceleration tools to access such platforms (5.25\% vs. 36.69\%).

To evaluate correlation between the use of international gaming platforms and the use of a ladder or booster, Chi-squared tests with an alpha level of 0.05 were used on data from online gamers who successfully evaded the APS. The analysis reveals statistically significant correlation ($p =  3.36\mathrm{e}{-63}$, thus smaller than $0.001$). Separate Chi-square tests only for male and female participants produced similar results, i.e., $p = 1.66\mathrm{e}{-18}$ for males, $p = 1.66\mathrm{e}{-31}$ for females.

Of the 279 respondents who used GFW evasion or game acceleration tools, 85 (43 male, 34 female, 8 undeclared) did not play games on international platforms, and are perhaps more likely to have used a ladder to evade the GFW.

The 2,277 survey respondents who were 19 or older, thus too old to have been affected by the third generation AAP, were further compared against the 1,087 that were 18 during the survey, thus might have been affected by them. Figure~\ref{fig:survey:gen3} reveals a substantial increase in the 18 year old group in the percentage of those affected by anti-addiction policies (71.30\% vs. 44.40\%) and slight increases in those that evaded addiction prevention systems (20.70\% vs. 19.72\%), played on international platforms (12.24\% vs 10.54\%) and used GFW ladders or accelerators for this (11.75\% vs. 9.02\%). Chi-square tests with Bonferroni correction revealed that the difference in perceived impact of the policies is statistically significant. However, the differences between the two groups on APS evasion, use of international platforms and use of evasion tools, are not statistically significant.

While the survey was conducted only months after the third generation policies were introduced, these numbers suggest the potential of policy changes to impact young gamers and their evasion behaviors. The following ($\S$~\ref{sec:aps:perceptions} - $\S$~\ref{sec:findings:gfw}) discusses qualitative study findings on the research questions of $\S$~\ref{sec:introduction}.

\section{RQ1: APS Perceptions}
\label{sec:aps:perceptions}

This section first details participant demographics, then discusses interview participant perceptions of anti-addiction policies, and suggested improvements to AAPs and APSes.

\begin{figure}
\centering
\includegraphics[width=\columnwidth]{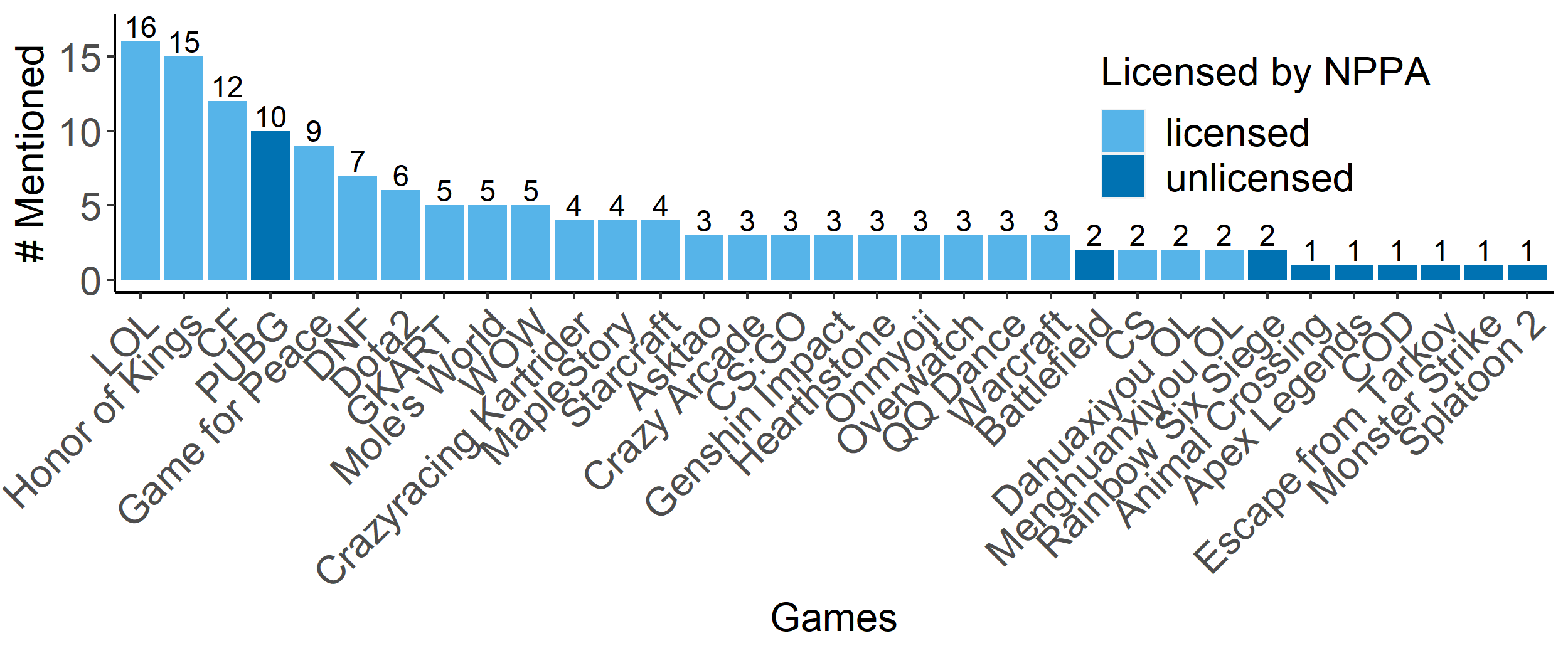}
\vspace{-10pt}
\caption{Online games played by interview participants while minors. With notable exceptions, most mentioned games are licensed by the NPPA.}
\vspace{-15pt}
\label{fig:games:counts}
\end{figure}

\noindent
{\bf Demographics}.
The 35 interview participants (22 male, 13 female, aged 18 to 28) with an educated background include students (thirteen undergrads, six MS, and two PhDs), technical college students (nine), or employed (three). Two participants started playing online games in kindergarten, nineteen in primary school, and nine in middle school. All interview participants used to play online games on laptops or PCs; 29 also played on mobile devices, and two on game consoles.

The participants' age, their age when they started playing online games, and the AAP timeline ($\S$~\ref{sec:background:timeline}) suggest that all participants started playing online games during the first generation AAP. Sixteen participants were only affected by the first generation. Nineteen were also affected by the second generation; Of these, seven might have also been affected by the third generation.

\noindent
{\bf Consistency Checks}.
All interview participants provided answers consistent with their earlier survey responses, i.e., (1) interview question I1 ($\S$~\ref{appendix:interview}) and survey assertion S2 ($\S$~\ref{appendix:survey}) on playing online games while being minor, (2) I4 vs. S7 on playing international games, (3) I7 vs. S4 on being affected by APSes, (4) I9 vs. S5 on evading APSes, and (5) I20 and I21 vs. S8 on using tools to evade restrictions. Further, all interview participants provided consistent answers to questions I5 and I17, on accessing non-gaming international sites.

\subsection{Impact and Perceptions of APS}

All 35 interview participants confirmed to have been affected by anti-addiction policies. They reported limits on gaming time and purchases, consistent with the AAP generation experienced ($\S$~\ref{sec:background}). Several participants confirmed NPPA's licensing process ($\S$~\ref{sec:background}), e.g. {\it ``The characters cannot be skeletons or have sexy outfits. Bloody and violent content is prohibited. These types of games do not get permission.''} (P68).

Figure~\ref{fig:games:counts} shows the NPPA-licensed vs. unlicensed games played by interview participants. Around 15\% (11 out of 53) of the online games played by participants were not NPPA-licensed; 18 of the 35 interview participants played at least one unlicensed game (on alternative platforms, see $\S$~\ref{sec:evasion:aps:alternative}).

\begin{figure}
\centering
\includegraphics[width=\columnwidth]{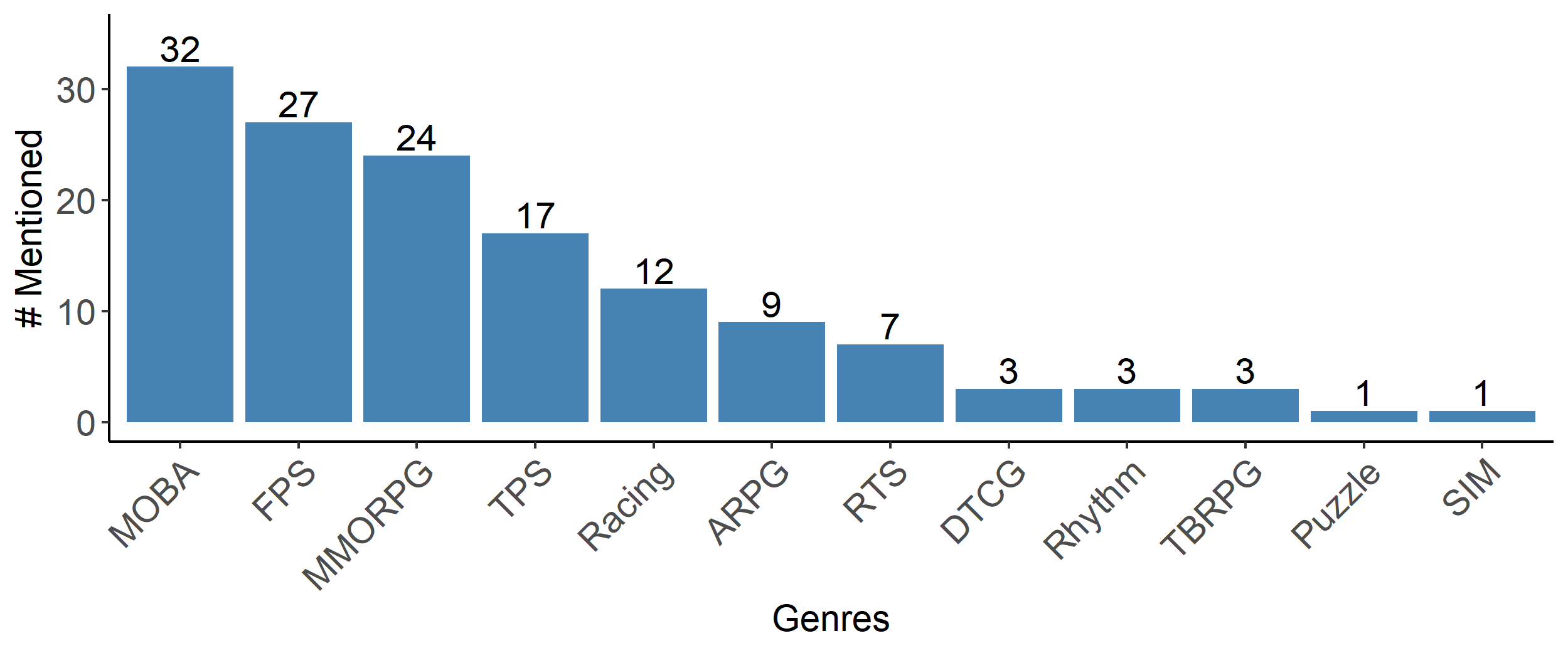}
\vspace{-10pt}
\caption{Game genres mentioned by interview participants.}
\vspace{-5pt}
\label{fig:genres:counts}
\end{figure}

Figure~\ref{fig:genres:counts} shows the distribution of the genres of online video games mentioned by participants.
Multiplayer Online Battle Arena (MOBA), First-person Shooter (FPS), Massively Multiplayer Online Role-playing Games (MMORPG), Third-person Shooter (TPS), and Racing were each mentioned by more than ten participants. Action Role-playing Games (ARPG), Real-time strategy (RTS), Turn-based Role-playing Games (TBRPG), Digital Collectible Card Games (DTCG), and Simulation (SIM) games were less popular.

Thirteen participants were critical of APSes even after no longer being affected by them. Five of them believed that AAPs are not a good idea, while the other eight criticized them even though they said they support the anti-addiction mission. Some perceived them to be unnecessary, e.g., {\it ``I do not like them because people in other countries do not have this issue''} (P155), and considered them to be too restrictive and inflexible: {\it ``The {\bf government should not have this one-rule-fit-all policy} without any flexibility''} (P124), {\it ``the APS does not respect the students' feelings, it is too heavy''} (P100).

Five participants were critical of the latest policies, e.g., {\it ``the previous AAP was acceptable, but the latest version is overkill''} (P680),
%
%
{\it ``the latest APS is too strict, even though I have not been affected by it.''} (P1382).

However, 15 participants were supportive of APSes and even considered them to be necessary, e.g., {\it ``the APS is a good addiction prevention method for minors.''} (P676). The policies were perceived to help people focus on their studies, have time for meaningful in-person activities, and avoid debts, e.g., {\it ``playing games affected my studies. The AAPs at least controlled my gaming hours and prevented me from addiction''} (P680). Of these, seven participants considered the policies to be too relaxed when they experienced them. Their criticism revolved around the lack of effectiveness of policy implementations, {\it ``The restrictions should be stricter. There are too many loopholes''} (P262), {\it ``many game providers are just pretending to enforce the policies''} (P30).
%

Two participants had attitudes toward AAPs that changed after no longer being affected by them, e.g., {\it ``when I was young I thought it was too strict, but now I think that the time limits are beneficial for minors''} (P110).
%

In summary, some of the participants who were affected by addiction prevention systems now consider them too relaxed. Others were critical even though they supported them. These polarized and conflicting perceptions are consistent with previous findings regarding support of censorship among Chinese Internet users~\cite{KSN17, WM15}: some people support censorship efforts even though they evade the GFW.

\subsection{Suggested APS Improvements}
\label{sec:aps:suggestions}

Participants suggested several APS improvements:

\noindent
{\bf Parental Control}.
Four participants suggested more influence for parental control, e.g., {\it ``it is better to let parents set limits''} (P8). Some discussed the importance of education, {\it game companies can add parent instruction lessons before their kids start playing games, to educate them on how to guide their kids scientifically''} (P124). Platforms like NetEase have indeed designed controls for parents to monitor and limit game hours and purchases~\cite{NetEase18}. However, several participants reported opening accounts on behalf of adult family members, in order to evade the APS ($\S$~\ref{sec:evasion:aps:id}). Family members are occasionally complicit, by going as far as helping their children bypass face recognition; this however provides parents with more opportunities to decide game types and times.

However, previous research suggests that restrictive rules and regulations set by parents on the child’s video-gaming behaviors are not effective in reducing pathological symptoms of video-gaming~\cite{CSLGK15}. Case in point, the adults that lend their IDs to minors in their family, may not know that they are used for online gaming: most online services in China require users to link IDs and use a phone during login, e.g., {\it ``Many of my classmates, whose parents work in other cities, live with their grandparents. The elderly do not know what is the APS. My classmates can easily get their IDs''} (P1476).


\noindent
{\bf Reversal to Earlier APS Generation}.
Several participants suggested reinstating rules from the 2nd generation AAP, e.g., {\it I would change the APS to the previous version, which allowed minors to play two hours a day} (P1382)
%
%
Some suggested age-based limits consistent with the 2nd generation AAP, e.g., {\it ``time limits should be different for different age ranges. Older minors, such as 16-18 years old, should have longer game hours than the younger ones''} (P192).

\noindent
{\bf Improved Real-Name Verification}.
In contrast with the previous suggestions, three participants suggested improvements to the policy implementations, including more effective verification methods, e.g., {\it ``the current system needs extra face recognition verification''} (P171), {\it ``I heard that you can use a photo to bypass the facial recognition verification system. And people can still use fake IDs to sign up for accounts''} (P60), and cross-platform limitations, e.g., {\it ``all the games should share the same time limitation, otherwise, gamers can play one hour for each different game''} (P64). We note that this stance is consistent with the earlier-reported finding that some participants wish for more rigorous APSes ($\S$~\ref{sec:aps:perceptions}).

\section{RQ2: APS Evasion}
\label{sec:evasion:aps}

Even though 18 interview participants were supportive of the anti-addiction policies, all 35 participants evaded addiction prevention systems. Many of them claimed that the APS implementations are ineffective and can be evaded. Consistent with previous remarks that gaming platforms are not doing enough, several participants explained that efforts to evade these policies, succeed because platforms need to be profitable, e.g., {\it ``face recognition, real-name verification, SMS verification, are very mature now, but some gaming platforms deliberately set up backdoors to allow minors to play.''} (P30).

To investigate such backdoors and RQ2 (a), we asked interview participants to describe their evasion experiences. In the following, we discuss our findings, structured according to the policy components illustrated in Figure~\ref{fig:aps:components}.

\subsection{Adult Accounts}
\label{sec:evasion:aps:accounts}

Participants revealed strategies to acquire adult accounts, for which anti-addiction policies do not apply.

\noindent
{\bf Account Rental}.
To evade the real-name verification component of the APS (Figure~\ref{fig:aps:components}), fourteen participants relied on China's market for renting gaming accounts. Since rented accounts are registered to real, adult users, they enable buyers to evade the APS. Account rental services are easy to access via search engines and e-commerce platforms, such as Taobao~\cite{taobao}, and are extremely affordable even for minors, e.g., 2-3 CNY (\$0.5) per hour~\cite{TencentRent}.

For convenience, some game account rental service providers, e.g.,~\cite{TencentRent}, have their own PC and mobile applications. Participants described the protocol for using such account-renting apps. First, they search for the game in the app, they pay for the service, e.g., using WeChat or Alipay. After switching to the game, the app auto-fills the rented account name and password, and the verification code received on the account owner's phone. The app interrupts the session when the rental time expires.

Participants explained that account rental services work because gaming platforms allow them, e.g., {\it ``{\bf most game companies don't apply very strict IP or location verification strategies}''} (P216). This could be because platforms are turning a blind eye ($\S$~\ref{sec:aps:perceptions}), or because gaming accounts can be rented for valid reasons, e.g., improving and diversifying gaming experience, game companions~\cite{GameCompanion}~\footnote{Game companions are hired to accompany users through the game.}, and gaming trainers, e.g., {\it ``There are some people who work as trainers, they need to rent game accounts to run their business''} (P110).

\noindent
{\bf Account Purchasing}.
Buying adult gaming accounts can also be used to evade the addiction prevention system. Buyers can find sellers on third-party account trading platforms, or even on social networks. Two participants claimed that account trading is allowed by game providers, e.g., {\it ``For Fantasy Westward Journey, the website [redacted] allows players to trade their game accounts. The owner needs to unlink their phone number and ID, then link to the new owner''} (P110).
%

Some participants explained that trading accounts will not allow minors to avoid the APS, e.g., {\it ``it should not be able to help you bypass the APS, because the account you bought needs to be linked to your ID''} (P168), {\it ``trading accounts requires linking your ID and mobile phone number, and after linking to a minor ID, they are still limited''} (P95). Others however revealed that linking a new ID is not necessary, and the buyer only needs to transfer the account to a new SIM card device, e.g., {\it ``log in to the account you purchased, and ask to transfer it to a new device. Then, an SMS is sent to the device of seller, who needs to confirm it and provide the phone number of the new device. A new SMS code is sent to the new device to finish up the binding''} (P155).

Four participants purchased accounts on international game platforms, e.g., {\it ``I can buy overseas Steam accounts on Taobao. It is cheaper to buy games from accounts registered in Argentina, Russia, Turkey''} (P142).
%

\noindent
{\bf Account Borrowing}.
\label{sec:evasion:aps:borrow}
Thirteen participants borrowed accounts from adult friends and family, e.g., {\it ``When one of my friends was not playing the game, {\bf I would type in their account name and password, then they would forward the 2FA code from their phone}''} (P9). Some also borrowed devices from adults, with access to adult gaming accounts, e.g., {\it ``my older sister used to ask me to help her play online games, she just gave me her phone directly''} (P81).
%
%
We discuss parental control in Section~\ref{sec:aps:suggestions}.

\subsection{Adult IDs}
\label{sec:evasion:aps:id}

National IDs are required to open online game accounts. They allow gaming platforms to verify the name and age of their account owners, which is a key APS feature. The Chinese national ID number consists of a 17-digit {\it body code} and a one-digit {\it verification code}. The body code consists of (from left to right): a six-digit address code, eight-digit date of birth code, and a three-digit sequence code. The following details participant-revealed sources of national IDs.

\noindent
{\bf Family IDs}.
Twelve participants used to open online game accounts with IDs from adult family members. Two did it without consent, {\it ``{\bf I used my parents' ID cards secretly}, and later I used my adult friends' ID cards, who never registered for the same game''} (P155). Some family members provided their ID without knowing the purpose, {\it ``my parents always provided their ID to me and never asked the reason. {\bf They understand everything online requires to link an ID in China}, and they trust me, so they never bothered to ask''} (P124).

Five participants did it with the adult's consent, e.g., {\it ``I used my father’s ID to sign up for a LOL account, linked with my father’s phone. I needed to ask him to help me log in when I wanted to play} (P87)''. Some asked family members to also help them bypass face recognition {\it ``Grandparents are happy to help bypass facial recognition''} (P1476). This suggests ability for family members to restrict the time spent by minors playing online games, see discussion in Section~\ref{sec:aps:suggestions}.

\noindent
{\bf Generated (Fake) IDs}.
To evade the Link ID component of the APS, (see Figure~\ref{fig:aps:components}), thus avoid providing their real, age-revealing ID, two participants 
%
%
mentioned creating adult gaming accounts using generated (name, ID) pairs, e.g., {\it ``I could create an account by looking for some ID number generators on the Internet''} (P262). Online ID generators provide not only the ID number, but also an associated random name and phone number, and matching DOB, gender, and address~\footnote{\url{https://www.dute.org/fake-id-card-number}}.
%
%
This worked because early APSes failed to validate IDs, e.g.,
%
%
{\it ``the ID number has specific patterns, and game providers didn't have an ID database''} (P30).

However, two participants mentioned that gaming platforms that implement the 3rd gen APS, are now able to detect generated IDs, e.g.,
%
%
{\it ``the fake names and ID cannot be used to register a new account. The game company has some way to check if they match''} (P171). Indeed, in addition to the government and police that maintain a database of valid IDs, other online sites, e.g.,~\cite{Tongchaba}, also provide paid ID verification services. While gaming platforms can use these services to identify fake IDs, we note that skilled users could also use them to find valid (name, ID) pairs, even for free. For instance, online train ticket purchasing services~\cite{TTC} require users to enter the ID and name at checkout time, and fail if the ID is invalid or does not correspond to the name.

\noindent
{\bf Purchased (Genuine) IDs}.
Another approach is to create adult gaming accounts using purchased genuine (name, ID) pairs online. Genuine person names and IDs are often sold online by their owners (voluntarily or forcibly)~\cite{SanHe}, or are stolen or scammed. Each ID on sale includes an ID number, the name of the owner, and photos that contain the face of the person to whom it belongs, and both sides of the ID card. Some ID sellers publish advertisements on blog sites such as tumblr.com. The sites selling IDs change frequently, and buyers need to contact them via QQ, e.g., {\it ``I can search for ID information online to create a gaming account, but links [to sites selling IDs] disappear later''} (P68).

\subsection{Alternative Gaming Platforms}
\label{sec:evasion:aps:alternative}

\begin{figure}
\centering
\includegraphics[width=\columnwidth]{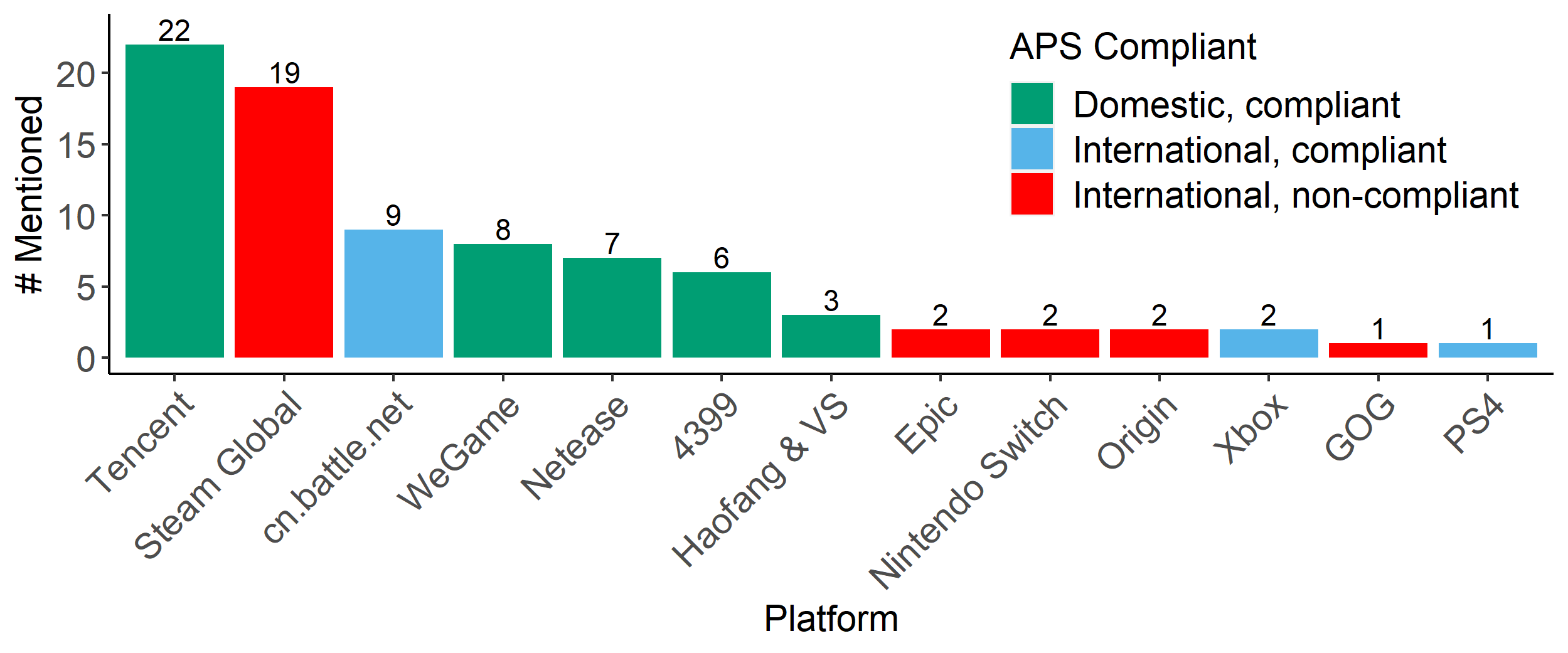}
\vspace{-20pt}
\caption{Gaming platforms mentioned in interviews, with associated mention counts. All domestic platforms are APS-compliant, and are most frequently mentioned. We observe however several popular non-compliant international platforms, notably Steam.}
\vspace{-5pt}
\label{fig:platforms:count}
\end{figure}

\noindent
{\bf International Gaming Platforms}.
More than 15\% (361) of the survey participants that played online games while minors, have used international gaming platforms. Figure~\ref{fig:platforms:count} shows the most popular platforms among our participants, and how many participants mentioned using them. Several such platforms are APS-compliant (Battle.net, Xbox, PlayStation), e.g., {\it ``Some overseas gaming platforms also require phone numbers for authentication''} (P216). However, other international platforms are non-compliant, e.g., {\it ``it is a gray area. These overseas game platforms always have a Chinese region option, but it is not regulated by the Chinese market''} (P168). The most popular such platform is the international Steam, which does not enforce AAPs on Chinese users.

Twenty-one interview participants confirmed playing on international platforms to avoid the APS, e.g., {\it ``I use international game platforms that are not affected by the APS, for instance, Steam, battle.net, Origin.''} (P35).
%
%
Participants explained that games on international gaming platforms impose higher latency, e.g., {\it ``there are restrictions on network speed''} (P216). In $\S$~\ref{sec:discussion:desensitisation} we discuss participant use of game accelerators and GFW ladders to alleviate this problem.

\noindent
{\bf Pirated Online Games and Servers}.
Several participants revealed a black market offering servers for pirated online games.
%
%
Operators often use private servers to run pirated versions of popular games, and create websites to advertise their servers~\cite{PiratedMMO19}. Such private gaming servers do not implement APS policies, e.g., {\it ``I used to play a pirated World of Warcraft game with my friends. It had no restrictions''} (P680), {\it ``I used to play Dota 1 on pirated Warcraft, so there were no restrictions for it''} (P262). Recent reports also mention pirates using Tencent's QQ accelerator app~\cite{TencentBooster} ($\S$~\ref{sec:discussion:desensitisation}) to sell access to Steam accounts ($\S$~\ref{sec:evasion:aps:accounts}) that owned pirated copies of popular games (e.g., ``Tale of Immortal'')~\cite{PiratedTencent}.

\subsection{Exposure to Fraud}
\label{sec:evasion:aps:exposure}

Participants revealed that efforts to evade the APS have exposed them to various types of fraud.

\noindent
{\bf Fake Gaming Platforms}.
Some participants reported that searches for international gaming platforms (Steam, Epic, Origin) also return links to fake gaming platforms. Participants explained that such platforms, e.g., \url{https://99box.com/}, {\it ``will obtain system permissions, then will popup ads windows on the desktop, even set the browser default page''} (P68). We confirmed these claims on a test machine, where we installed one such fake platform. Other participants reported fake platforms (e.g., \url{https://loginhell.com/login/steam}) that steal and resell user credentials, e.g., {\it ``once I logged in with my account information on the fake Steam game platform, and it stole my account password''} (P168). VirusTotal did not flag any of the fake platform URLs mentioned by participants.

\noindent
{\bf Fake Gaming Boosters}.
Attempts by online gamers to find and install game accelerators, expose them to (1) malware, e.g., {\it ``when I download some game boosters through the Internet and install them on my computer, it will also automatically install some malware and advertisement software''} (P676), and (2) malicious services, e.g., {\it ``A lot of game boosters are fake, they can not be used after installation. Once I logged in with my account information, it directly stole my account password''} (P168). Some participants said that such phishing attacks take place over the chat of platforms they access, e.g., {\it ``some ads in [in-game] chat channels claim they can help gamers evade the APS system. They will provide a phishing web page, and once you type your account and password, they steal your in-game items and even your accounts''} (P110).

While some gaming platforms attempt to prevent such attacks, e.g., by disallowing links, filtering posts that contain certain keywords, attackers find ways around them, e.g.,  {\it ``fake ads use alternative text characters to cheat the keyword filtering system. For example, WeChat is wx, AliPay is zfb''} (P110). This is consistent with other reports of Chinese users using substitute terms for blocked keywords~\cite{M14, KKGB17}.

\noindent
{\bf Account Purchasing Scams}.
Participants revealed experiencing scams when purchasing gaming accounts, e.g., {\it ``some scammers publish ads selling adult accounts, claiming they are real-named verified. However, they disappeared without delivering the valid accounts after collecting the money''} (P1382). Others revealed that the scam takes place {\it after} purchasing the account, e.g., {\it ``Someone sold their account, then used their linked ID information to reclaim the account via the game platform account appealing process''} (P171). Some were not aware this is a scam, e.g., {\it ``I used to purchase accounts, but since they are not binding to any real person, these accounts would be eventually deleted by the game company''} (P60). Other participants however revealed the necessity to transfer ownership (or at least the registered phone number) when purchasing an account ($\S$~\ref{sec:evasion:aps:accounts}).

\section{RQ3: Experience and Evasion of GFW}
\label{sec:findings:gfw}

All the interview participants were aware of the GFW and had heard about GFW evasion. Twenty-two of them mentioned that they were affected by the GFW, and also perceived that the GFW affects their ability to play online games, e.g., {\it ``I cannot play on international game platforms without evading the GFW''} (P68)
%
%
Tens of millions of Chinese users have accounts on international gaming platforms~\cite{ChineseGameMarket, SteamDark}. While the GFW does not block access to these platforms, it blocks some of the pages on these platforms that are used by gamers to communicate and exchange content, e.g., the Steam community and workshop~\cite{SteamBlocked}.

\begin{figure}
\centering
\includegraphics[width=\columnwidth]{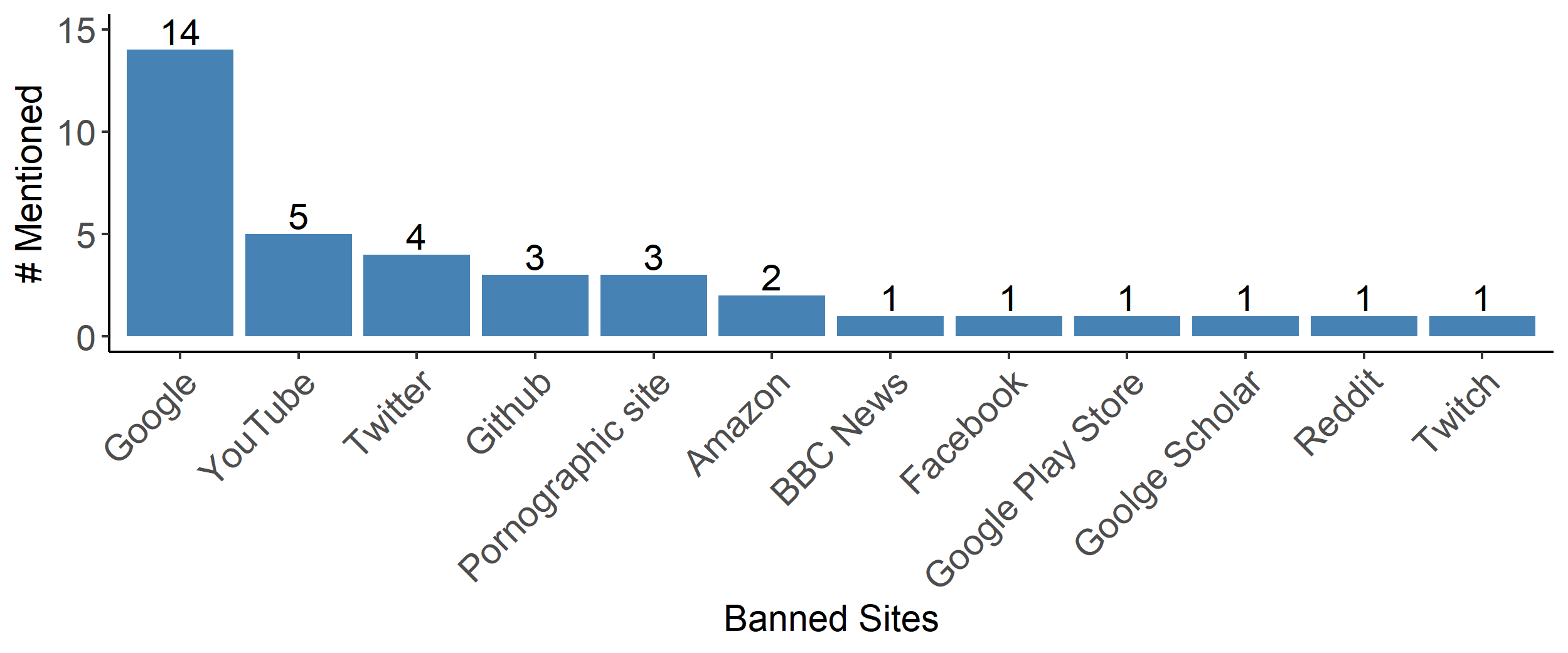}
\vspace{-10pt}
\caption{Censored sites that interview participants access once they evade the GFW.}
\vspace{-15pt}
\label{fig:sites:counts}
\end{figure}

Fourteen participants had evaded the GFW and were open to discussing their experiences. Several participants used the term VPN to refer to any GFW-evasion tool, including those that are not VPNs. Reasons mentioned by participants to evade the GFW include playing online games, streaming gaming videos, accessing search engines, social networks, online forums and shows, tools and apps, academic papers, IT courses, study material, porn, and news. Figure~\ref{fig:sites:counts} lists the most popular foreign, non-gaming sites accessed by interview participants after evading the GFW. Notably, two participants evade the GFW to access diverse opinions, e.g., {\it ``{\bf Some people think censorship is mind control}. I don’t buy it, so I evade the GFW to verify information from both sides. I can check if different parties are reporting the truth'' (P110)}. 

Participants explained that their GFW evasion sessions last between 30 mins to 2 hours. Most claimed daily evasion, while a few only do it a few times per week or month.

To determine if participants report experiences from current vs. past GFW-evasion activities, we asked those who were comfortable discussing GFW experiences and evaded the GFW, to read the latest post (6 digit code) of a Twitter account controlled by an author. Twelve of the 35 interview participants provided this information and did so in a few seconds. This suggests that these participants were evading the GFW at the time of the interview. Since the Twitter account and post are harmless, this activity did not expose the participants to risks higher than those they already incur in their regular GFW-evading activities. One limitation is that not accessing the code during the interview does not imply that the participant is not currently evading the GFW.

\subsection{GFW Ladder Discovery and Payment}

Most participants find GFW evasion tools online. Several participants explained that recently, (domestic) search engines did not produce working solutions, e.g., {\it ``I used to search for it, but now I can't find anything that works} (P60). To confirm, we searched on domestic search engines (e.g., Baidu) using ``VPN'' (``\begin{CJK*}{UTF8}{gbsn}翻墙\end{CJK*}''), and obtained no valid results. However, participants reported that ladders can still be found when searching for the real name of the ladder, or Chinese jargon and homonyms, e.g., {\it ``first, I use homonyms or special terms to search for the ladder tutorial, and then I can reach the ladder through the link in the tutorial or by directly searching the ladder name mentioned in the tutorial''} (P1599).

Others find such tools from (1) friends and classmates, e.g., ``{\it College mates, we tell each other the information''} (P110), (3) the app store, e.g., {\it ``there was an app with a list of tools to evade the GFW''} (P9), (3) online forums (e.g., Tieba~\cite{baidutieba}) and group chats on QQ, and (4) other people in internet cafes.

Several participants explained that they started with a free VPN to evade the GFW, then they were able to easily find a better tool, e.g., {\it ``I found tutorials in Baidu Tieba. After evading the GFW, I saw other solutions promoted on YouTube videos. The videos will have VPN advertisements, and YouTubers will also promote that a particular VPN is efficient''} (P155). Nine of the eighteen GFW ladders mentioned by participants, are either free or have a free tier.

Sixteen of the eighteen GFW ladders mentioned by participants, were available for download from China, following searches that included jargon, and homonyms, e.g., ``\begin{CJK*}{UTF8}{gbsn}蕃蔷\end{CJK*}'' (homonym of climbing the wall). Major foreign VPN providers like Express VPN and Nord VPN are no longer effective in China, since they currently have no VPN server nodes available in mainland China.

\noindent
{\bf Payments}.
Most participants explained that the VPNs that they used were accepting the Chinese UnionPay system, WeChatPay or AliPay, e.g., {\it ``the VPN providers charge via the payment code of Ali or Tencent''} (P30). One participant (P416) mentioned Paypal and international credit cards. Indeed, 12 of the 18 GFW ladders mentioned by participants, accept Chinese payments. While the foreign-operated Express VPN and Nord VPN used to provide Chinese-inclusive payment methods, they no longer seem to do so.

\subsection{Reliability}

A widespread theme among our participants is the unreliability of GFW evasion techniques, in particular VPN services, e.g., {\it ``There are many VPNs that are very unstable once you pay for them''} (P95). Even VPNs that work, fail frequently, e.g.,
%
%
{\it ``The VPN will go down randomly, and for a while I cannot use it at all.''} (P87). Some evasion tools are reported to fail at the same time, e.g., {\it Most of the time, {\bf when one VPN goes down, others will not work either}. But after a while, they start working again''} (P87).

Previous work has shown inconsistencies in the operation of the GFW~\cite{EWMC15}. Several participants remarked that unreliability is due to GFW efforts, e.g., {\it ``some ladders are often banned.''} (P9), {\it ``many free VPNs are blocked soon after being released''} (P262). This is consistent with a race between the GFW and ladders, where the GFW discovers and blocks the IPs of their servers, while ladders acquire new ones.

Three participants perceived reliability to be an essential feature of GFW ladders, e.g. {\it ``It is quite troublesome to research and configure a new VPN each time, it would be great if the service was stable''} (P1599). Others however were less concerned, perhaps due to the normalization of frequent failures of GFW evasion tools and the development of contingency plans, e.g., {\it ``{\bf There is an app that gives you a list of all the wall ladders}. Once one ladder stops working, I download another via this app''} (P168).

\subsection{Privacy Concerns}

Of the 18 GFW ladders revealed by participants, 16, including some of the free ones, require users to create accounts. Further, some VPNs, including Facebook's Onavo Protect~\cite{OnavoProtect}, have been shown to log data from their users~\cite{SensorTower}. This is expected from free VPNs, since if the user is not paying for the product, they are likely the product~\cite{OnavoProtect}. Previous work has also shown that some VPN apps do not encrypt the tunneled traffic~\cite{IVSKP16, KDVSKV18}, and many misrepresent the physical location of their vantage points, often situated in countries other than those they advertise~\cite{KDVSKV18}. In addition, some VPNs are suspected of being owned by Chinese companies~\cite{VPNOwnership}, and may be compelled to share data with the censor.

Half of the respondents perceived privacy to be an essential feature offered by GFW-evasion tools. When asked if they had privacy concerns related to their VPN provider, e.g., logging and sharing their traffic with others, two participants were concerned about exposing login, IP address, and payment information, e.g., {\it ``{\bf I worried that the GFW ladder stores my IP information}''} (P262). However, seven participants were not concerned about being monitored by GFW ladders. One reason is that they find these tools after evading the GFW, and they pay for them, e.g., {\it ``VPN providers prefer to earn money from their tools, and try to provide more secure services, rather than hack me''} (P110). Another reason is that some VPNs seek their users' collaboration, e.g., {\it some VPNs will send a pop up to notify me to stop browsing specific websites when the traffic is under inspection''} (P110).

Two participants have no concerns related to government monitoring of their use of GFW ladders,
%
%
``{\it
%
%
There is nothing to worry about, if you don't do illegal things. No one cares if you only watch porn sites but don't spread the content''} (P1599). However, participants who reported official, free VPN services to allow students and researchers access to websites of foreign universities, academic platforms, and study materials, are worried about being monitored while using these VPNs, e.g., {\it ``I would not watch porn via this channel, since {\bf access is definitely monitored}''} (P9).

\subsection{Exposure to Malware and Illegal Content}

Five participants explained that searching, installing and using GFW evasion tools exposes them to malicious software, e.g., {\it ``When I launch some free ladder, it will popup ads and automatically download spam software''} (P171).
%
%
This is consistent with previous work that has shown that over 38\% of Android apps that use the Android VPN permission have some malware presence~\cite{IVSKP16}.

Ladders also expose users to illegal ads (e.g., gambling, porn, money lending), both during search, e.g., {\it ``When I was searching for VPNs, a lot of illegal ads popped up''} (P124), and during use, {\it ``Half of the ladders will show illegal ads, and some ad links are fake''} (P171). Previous work has found free VPN apps that inject JavaScript codes for advertising purposes~\cite{IVSKP16}.

We note that such problems will likely persist, since they are compounded by the sensitivity of the topic: victims may find it difficult to report malicious VPNs to the authorities.

\section{Discussion}
\label{sec:discussion}

This section discusses the future of addiction prevention systems, the potential of anti-addiction policies to normalize censorship evasion, and study limitations.


\subsection{The Future of APS}
\label{sec:discussion:future}

The survey revealed significant impact of third generation anti-addiction policies: 71.30\% of the 18 year old respondents reported being affected by anti-addiction policies when compared to 44.35\% among those that were not affected by the third generation policies. This is likely to intensify. For instance, in the post 3rd gen AAP era, NRTA (sibling department of NPPA) released a regulation mandating all live streaming platforms prevent users from live-streaming unlicensed video games~\cite{NRTA22}. Tencent also 
discontinued its gaming accelerator service, which allowed Chinese users to access overseas games~\cite{TencentBoosterStop}. Further, almost all online game companies and platforms in China have signed a convention to prevent minors from renting gaming accounts~\cite{ConventionTencent21}.

Addiction prevention systems may also evolve, e.g, as proposed by several interview participants, to implement more effective identity checks ($\S$~\ref{sec:aps:suggestions}). While Tencent's mobile gaming platform has implemented live facial recognition of its users~\cite{Tencent18} ($\S$~\ref{sec:aps:components}), some account rental services provide facial validation services, where the account owner is online to assist the renter. Future APSes may further compare the IP addresses used during normal gaming and during identity checks. Further, they could require users to keep cameras always on, to conduct surreptitious identity checks. However, we also observe conflicting requirements for gaming platforms, due to their high revenues from minors.

\subsection{Desensitisation of Evasion}
\label{sec:discussion:desensitisation}

\noindent
{\bf Relation Between GFW and APS Evasion}.
Section~\ref{sec:evasion:aps} discussed the use of international gaming platforms to evade the APS, and resulting latency issues. To address this problem, 17 interview participants used game accelerators (boosters) or GFW ladders, to connect to international platforms.
%
%
%
A few participants reported that evasion of the GFW is not necessary to access international gaming platforms.
%
%
This is because even though many games (including foreign ones) are not licensed in China ~\cite{BannedGames}, most international platforms are not blocked by the GFW. The GFW only blocks pages on gaming platforms that are used to communicate and exchange content, e.g., the Steam community and workshop~\cite{SteamBlocked}.


However, 17 participants who evaded the GFW revealed that their reasons include, among others, playing and watching games online. In addition, several participants use the same tool to boost games and to evade the GFW
%
%
Some participants also confused gaming accelerators and GFW evasion tools (e.g., mentioning use of ladders to access games, and accelerators to evade the GFW). Perhaps one reason for this confusion is that several popular VPNs also have game acceleration functionality~\cite{PandaVPNBooster, ExpressVPNBooster}.

All the GFW ladders mentioned by participants can also be used for game acceleration purposes. Further, some accelerators (e.g., Steam++~\cite{SteamTools}) also provide partial GFW-evasion capabilities, e.g., enabling access to some censored sites (e.g., GitHub, Google Authenticator, Pixiv, Discord, Twitch).

\noindent
{\bf Evasion Desensitisation Hypothesis}.
The user confusion and functionality overlap between GFW and APS evasion tools suggest an intersection between efforts to evade the GFW and APS: some participants use VPNs to evade both the GFW and the APS, while game boosters provide ability to evade both the APS and partially the GFW. We conjecture therefore that online gaming restrictions have the potential to normalize the use of GFW-evading technologies by Chinese minors, particularly in light of several participants starting to use such tools from an early age.

We observe relationship to previous work that argues that normalization of censorship increases with exposure~\cite{WM15}, and that censorship of non-politically threatening topics leads to normalization of censorship in China~\cite{Y21}. However, going in the opposite direction, and building on the desensitization theory~\cite{CAB07, FVHA09}, we posit that the need to evade anti-addiction policies and the GFW for trivial activities, e.g., playing games online and accessing non-threatening but umbrella-censored sites, may desensitize young Chinese gamers to the fear that evading the GFW will violate the law. Case in point, 15 of 35 online gaming participants use GFW evasion tools, and were willing to discuss their evasion experiences with us.

Desensitization may occur when users are unaware of the implications of their actions, e.g., switching from using VPNs or gaming accelerators for gaming, to Internet browsing without restrictions. Then access to the unrestricted Internet may become the new norm, and may further encourage exploration of the GFW-censored web. This is perhaps related to {\it normative dissociation} experiences, characterized by absorption, diminished self-awareness and often a reduced sense of time and control, and a gap in memory~\cite{B06}. Indeed, some participants in Baughan et al.~\cite{BZRLSBH22}'s study report passively slipping into normative dissociation while using social media.


We acknowledge however that desensitization is not a theme that emerged from our study, but a hypothesis. Future work may further investigate its validity.

\subsection{Recommendations for CRS Developers}
\label{sec:discussion:crs}

The popularity of game accelerators can be used by their users to plausibly explain GFW evasion and the presence of evasion-enabling software on their devices. Further, a strategy for developers of censorship resistant systems (CRSes) to increase adoption among young Internet users in China and provide them with some measure of plausible deniability if their devices are inspected by the censor, is to market them as gaming accelerators. However, several participants observed that censorship evasion tools that become popular, also attract attention from the censor, which often leads to banning efforts. Thus, CRSes could provision mechanisms to limit their perception of popularity, for instance through re-branding (e.g., open source Steam++ accelerator is now Watt Toolkit), or capping the number of active accounts.

\vspace{-5pt}

\subsection{Limitations}
\label{sec:limitations}

\vspace{-5pt}


While we received 3,364 valid data points, including 77 non-students (both unemployed and employed), and interviewed four non-students, our sampling recruitment procedure is biased toward a younger and better educated population. However, evasion of the GFW and of gaming restrictions is also more likely to occur among this demographic.


%

\noindent
{\bf Novelty of Revealed Circumvention Methods}.
A seeming limitation of this work is that knowledge of the methods revealed by study participants is readily available to Chinese users via e-commerce sites, search engines, game forums, friends, and family. However, one goal of this paper is to expose the research community and CRS developers to anti-addiction and GFW circumvention strategies popular among young and educated Chinese netizens.

\noindent
{\bf Limited Investigation of 3rd Gen AAP}.
The studies were conducted shortly after the 3rd generation AAP came into effect (September 2021). Thus, participant testimonies provide little perspective on their effects. However, restriction mechanisms, goals, enforcement actions, and agent behaviors, evolve almost monthly. Any study will intrinsically present a snapshot of only the present and past. This work is a snapshot of the present in the context of the past over two decades of gaming restrictions (first policies were introduced in 2000). Follow-on studies may explore future changes.

Further, the latest AAP generation might have impacted partnerships between international game designers, e.g., Activision Blizzard, and Chinese gaming platforms, e.g., NetEase (see $\S$~\ref{sec:background}), leading to some international games not being released in China. Our study does not investigate the indirect impact of the AAP, e.g., through recently unlicensed games, on the interest of Chinese netizens in circumvention tools. We note however that new games are released almost weekly, including in the post third-gen AAP period. Importantly, of the subset of games not licensed and operated in China, many are not GFW-blocked, and are easily accessible without ladders. This includes Activision Blizzard games.


\noindent
{\bf Survey Ecological Validity}.
Survey participants were free to stop the survey at any time and could skip any question. While this may limit the ecological validity of the survey, only 18 participants who started the survey did not complete it. Further, all of the 3,364 respondents whose data was reported, chose an answer (Yes or No) for the assertion about playing online games. All of the 2,415 respondents who answered in the affirmative, also chose an answer for the assertions about being affected by the APS and evading the APS.

\noindent
{\bf Demographics Limitation}.
The gender question in the survey only provided ``Male'' (\begin{CJK*}{UTF8}{gbsn}男\end{CJK*}), ``Female'' (\begin{CJK*}{UTF8}{gbsn}女\end{CJK*}) and ``Prefer not to answer'' (\begin{CJK*}{UTF8}{gbsn}不愿回答\end{CJK*}) options. Thus, participants who do not identify as male or female might have either self-described as binary, or might have chosen the ``Prefer not to answer'' option. In the open-ended gender interview question, all participants self-described as binary.

\vspace{-5pt}

\section{Conclusions and Future Work}
\label{sec:conclusions}

\vspace{-5pt}

This paper explores the intersection between the APS and GFW evasion ecosystems and concludes that anti-addiction policies do not work as designed against even very young gamers. Over 27\% of the surveyed online gamers were able to evade addiction prevention systems, while being minors. Somewhat counter-intuitively, APSes seem to act as a general censorship evasion training ground for tomorrow’s adults, by familiarization with and normalization of general evasion techniques, and desensitization to their dangers. Paper findings may help CRS developers to design tools that leverage technologies, services and platforms popular among the censored audience.

\vspace{-5pt}

\section{Acknowledgments}

\vspace{-5pt}

We thank the anonymous shepherd and reviewers for their insightful feedback and recommendations. This work was supported in part through NSF awards 2013671, 2052951 and 2114911.

\bibliographystyle{plain}
\bibliography{censorship,ethics,gaming,procedures,steganography,theory}

\appendix

\section{Informed Consent Excerpts}
\label{appendix:consent}

We include relevant information from the consent form.


\noindent
{\bf Risks and/or Discomforts}.
The study has the following possible risks to you: First, some of the questions that we will ask you may be upsetting. You can skip any questions you don’t want to answer, or stop the study entirely, at any time. Second, once we publish our results, other parties, including online game service providers, may try to close any loopholes in their defenses. If you are evading the great firewall you already incur this risk. However, since you are over 18 years old now, you are no longer impacted by the online game anti-addiction system.

%
%

\section{Survey}
\label{appendix:survey}


\noindent
S1. How old are you now?

\noindent
S2. I used to play online video games when I was underage.

\noindent
S3. I used to make purchases in online games when I was underage.

\noindent
S4. I was affected by the addiction prevention systems when I was underage.

\noindent
S5. I was able to evade the addiction prevention systems when I was underage.

\noindent
S6. I can understand foreign languages on sites outside China.

\noindent
S7. I used to play online games on gaming platforms hosted outside China.

\noindent
S8. I have used a GFW ``ladder'' or game booster to play online games on gaming platforms hosted outside China

\noindent
S9. What is your current occupation?

\noindent
S10. What is your gender?

\noindent
S11. Would you like to participate in a longer follow-up interview with similar questions on this topic? We will pay you 10 CNY (1.5 USD) for every 15 minutes that you spent with us in the interview, for instance, if you spent 30 minutes, we will pay you 20 CNY; 1 hour, we will pay you 40 CNY. Please type your preferred contact information.

\section{Interview}
\label{appendix:interview}

\noindent
I1. Did you play online video games when you were underage?
I1.b. How old were you when you started playing online video games?

\noindent
I2. Can you give me some examples of online games that you used to play when you were underage?


\noindent
I3. Can you give me some examples of online game platforms that you used when you were underage?


\noindent
I4. Did you play any non-Chinese online games? That is, games offered on sites that were hosted outside China.
I4(a) Did you have trouble connecting to such sites? For instance, due to the firewall?

\noindent
I5. What about other non-Chinese, non-game related sites?
I5(a). Are those sites in English?

\noindent
I6. Do any of the online games you used to play have the addiction prevention system implemented? 

\noindent
I7. Did the addiction prevention system affect you when you were a minor? How?

\noindent
I8. Do you think that the addiction prevention system is a good idea? Why?

\noindent
I9. Some people told us that they were able to evade the addiction prevention system. Did you ever try to evade the system when you were underage?
I9(a) Were you able to successfully evade the system?

\noindent
I10. I would like to discuss more about your experiences with evading addiction prevention systems. Do you feel comfortable talking about this? If the answer is yes, continue.

\noindent
I11. Some people shared some of their strategies to evade the addiction prevention system. Can you share some of the strategies that helped you evade the AA system?


\noindent
I12. How did you find the information about such evasion techniques?

\noindent
I13. Did you communicate with other players on any gaming platforms?

\noindent
I14. In your efforts to evade the addiction prevention system, did you ever encounter any attempts at fraud, for instance, sites or products that tried to profit from you?

\noindent
I15. I would like to discuss more about your experiences with the great firewall. Do you feel comfortable talking about this? If the answer is yes, continue.

\noindent
I16. Can you read and understand text written in a foreign language? Which ones?

\noindent
I17. Do you access any non-gaming sites outside China?

\noindent
I18. Were you ever affected by the Great Firewall? That is, have you ever tried to access any non-gaming sites banned by the GFW outside China?

\noindent
I19. Have you ever heard about GFW evasion?

\noindent
I20. Have you used any tools to try to evade the firewall? Did you succeed?

\noindent
I21. Why have you evaded the GFW?

\noindent
I22. How do you find GFW ladders?

\noindent
I23. Have you encountered any issues related to the GFW ladders that you use? 

\noindent
I24. Are you concerned about privacy issues when using GFW ladders?
For instance, have you ever been worried that the GFW ladder you use stores information about the sites that you access?

\noindent
I25. In your efforts to evade the GFW, did you ever encounter any attempts at fraud, for instance, sites or products that tried to profit from you?

%


\noindent
I26. How old are you?

\noindent
I27. What is your highest level of education?

\noindent
I28. What is your current occupation?

\end{document}